\documentclass[aps,pra,twocolumn,footinbib,floatfix,superscriptaddress]{revtex4-2}
\usepackage{graphicx}
\usepackage{indentfirst}
\usepackage{braket}
\usepackage{float}
\usepackage{amsmath}
\usepackage{physics}
\usepackage{amssymb}
\usepackage{CJK}
\usepackage{esint}
\usepackage{color}
\usepackage[T1]{fontenc}
\usepackage{subfigure}
\usepackage{amsfonts}
\usepackage{footmisc}
\usepackage{scrextend}
\usepackage{multirow}
\usepackage[outdir=./]{epstopdf}
\usepackage{svg}
\usepackage{algorithm}
\usepackage{algpseudocode}
\usepackage{soul}
\usepackage{mathtools}
\usepackage[english]{babel}
\usepackage{url}
\usepackage{comment}
\usepackage{array}
\usepackage{subfiles}
\usepackage{tikz}
\usetikzlibrary{shapes,arrows}
\usepackage{mathrsfs}
\usepackage[mathscr]{eucal}
\usepackage[hyperfootnotes=false]{hyperref}
\usepackage{cleveref}
\usepackage{stmaryrd}

\definecolor{darkblue}{rgb}{0,0,0.5}
\hypersetup{
colorlinks=true,
linkcolor=black,
filecolor=blue,
citecolor=darkblue,  
urlcolor=black,
}

\usepackage{trimclip}

\makeatletter
\DeclareRobustCommand{\shortto}{%
  \mathrel{\mathpalette\short@to\relax}%
}

\newcommand{\short@to}[2]{%
  \mkern2mu
  \clipbox{{.5\width} 0 0 0}{$\m@th#1\vphantom{+}{\shortrightarrow}$}%
  }
\makeatother

\newcommand\argmin{\mathop{\mathrm{argmin}}}

\usepackage{pict2e,picture,graphicx}

\makeatletter
\DeclareRobustCommand{\Arrow}[1][]{%
\check@mathfonts
\if\relax\detokenize{#1}\relax
\settowidth{\dimen@}{$\m@th\rightarrow$}%
\else
\setlength{\dimen@}{#1}%
\fi
\sbox\z@{\usefont{U}{lasy}{m}{n}\symbol{41}}%
\begin{picture}(\dimen@,\ht\z@)
\roundcap
\put(\dimexpr\dimen@-.7\wd\z@,0){\usebox\z@}
\put(0,\fontdimen22\textfont2){\line(1,0){\dimen@}}
\end{picture}%
}
\makeatother
\newcommand{\veryshortrightarrow}{\hspace{.2mm}\scalebox{.8}{\Arrow[.1cm]}\hspace{.2mm}}

\def\be{\begin{equation}}
\def\ee{\end{equation}}
\def\ba{\begin{eqnarray}}
\def\ea{\end{eqnarray}}
\def\bal{\begin{equation}\begin{aligned}}
\def\eal{\end{aligned}\end{equation}}

\def\bp{\begin{pmatrix}}
\def\ep{\end{pmatrix}}

\usepackage{bm}

\newcommand{\1}{^{(1)}}

\usepackage[normalem]{ulem}

\begin{document}

\title{Entanglement-assisted detection of fading targets via correlation-to-coherence conversion}

\author{Xin Chen}
\affiliation{
Department of Electrical and Computer Engineering, University of Arizona, Tucson, Arizona 85721, USA
}

\author{Quntao Zhuang}
\email{qzhuang@usc.edu}
\affiliation{
Ming Hsieh Department of Electrical and Computer Engineering, University of Southern California, Los
Angeles, California 90089, USA
}
\affiliation{
Department of Electrical and Computer Engineering, University of Arizona, Tucson, Arizona 85721, USA
}
\affiliation{
James C. Wyant College of Optical Sciences, University of Arizona, Tucson, Arizona 85721, USA
}
\begin{abstract}
Quantum illumination utilizes an entanglement-enhanced sensing system to outperform classical illumination in detecting a suspected target, despite the entanglement-breaking loss and noise. However, practical and optimal receiver design to fulfil the quantum advantage has been a long open problem. Recently, [arXiv:2207.06609] proposed the correlation-to-displacement (‘C$\veryshortrightarrow$D’) conversion module to enable an optimal receiver design that greatly reduces the complexity of the previous known optimal receiver [Phys. Rev. Lett. {\bf 118}, 040801 (2017)]. There, the analyses of the conversion module assume an ideal target with a known reflectivity and a fixed return phase.
In practical applications, however, targets often induce a random return phase; moreover, their reflectivities can have fluctuations obeying a Rayleigh-distribution.
In this work, we extend the analyses of the C$\veryshortrightarrow$D module to realistic targets and show that the entanglement advantage is maintained albeit reduced. In particular, the conversion module allows exact and efficient performance evaluation despite the non-Gaussian nature of the quantum channel involved.
\end{abstract}

\date{\today}

\maketitle

\section{Introduction}
Quantum entanglement enables performance boost in a wide range of optical sensing tasks, such as phase sensing~\cite{Escher_2011,gagatsos2017bounding}, target detection and ranging~\cite{Lloyd2008,tan2008quantum,zhuang2017optimum,zhuang2021quantum,zhuang2022ultimate}, loss sensing~\cite{sarovar2006optimal,venzl2007,monras2007,adesso2009,monras2010,monras2011,Nair_2011,nair2016,nair2018}, noise sensing~\cite{pirandola2017ultimate,shi2022ultimate} and gain sensing~\cite{nair2022optimal}. Despite the varieties of the applications, the sensing processes can often be modeled as bosonic Gaussian channels~\cite{weedbrook2012gaussian}, which preserve the Gaussian form of input Wigner functions. The Gaussian nature of the quantum channel enables efficient exact evaluation of the sensing precision, especially when the source is also Gaussian~\cite{Pirandola2008,banchi2020quantum}. Moreover, the structure of the Kraus operators of the bosonic Gaussian channel also allows the proof that Gaussian probes are optimal among all possible input states~\cite{Escher_2011,nair2020fundamental,nair2018,nair2022optimal,shi2022ultimate}.

Take target detection as an example, the transceiver-to-receiver path in presence of a distant target can be modeled as a Gaussian thermal-loss channel with low transmissivity; when the target is absent, the thermal-loss channel degrades to its zero transmissivity limit. In a quantum illumination (QI) protocol with the common Gaussian entangled source of two-mode squeezed vacuum, the error probability performance limit can be obtained via the efficiently calculable quantum Chernoff bound (QCB)~\cite{Audenaert2007,Pirandola2008}, which enables the surprising discovery of a six-decibel error exponent advantage over classical illumination (CI) despite loss and noise~\cite{tan2008quantum}. 

Things become challenging when non-Gaussian elements are inevitably involved. To begin with, although the channel and source are Gaussian, receivers based on only Gaussian operations (e.g., optical-parametric amplification and phase conjugation) are only able to achieve half of the error exponent advantage~\cite{Guha2009}. Previously proposed optimal receiver design relies on complex non-Gaussian operations that forbid exact performance evaluations~\cite{zhuang2017optimum,zhuang2017fading}. Moreover, a practical target detection scenario involves fading targets, where the random phase noise and fluctuating reflectivity make the quantum channel non-Gaussian. The non-Gaussian nature of the problem makes it difficult to evaluate entanglement's advantage in detecting fading targets.

%Moreover, the performance of Gaussian receivers degrades severely in presence of target fading~\cite{zhuang2017fading}.

%In particular, the fading quantum channel involved is also non-Gaussian, leading to further challenges in exact evaluation of entanglement advantage. Indeed, Ref.~\cite{zhuang2017fading} involves substantial approximations while the communication version of the problem entirely relies on non-tight lower and upper bounds~\cite{zhuang2021quantum-enabled}.

%In the case of target detection, only approximate results~\cite{zhuang2017fading} are available via the sum-frequency-generation receiver design~\cite{zhuang2017optimum}, while exact quantum advantages are unknown.

In this paper, we utilize the recently proposed correlation-to-displacement ('${\rm C}\veryshortrightarrow {\rm D}$') conversion module~\cite{shi2022} to evaluate entanglement's advantage in a practical QI target detection scenario with fading targets. The conversion module reduces multi-mode correlated state detection to single-mode coherent-state detection, enabling optimal receiver design and also efficient evaluation even when non-Gaussian elements are involved. Our results show that when there is only correlated phase noise across the probing, the error probability still decays exponentially with the number of probing. Entanglement's error-exponent advantage is still six-decibel when the signal brightness is extremely small, but degrades as the brightness increases. Such robustness resembles previous findings in the communication case~\cite{zhuang2021quantum-enabled}. In the presence of transmissivity fluctuation of the Rayleigh type, however, the error probability decays polynomially with the number of probing probes, and the advantage from entanglement is small, despite being non-zero.

\section{Model for fading target detection}

%
%\begin{figure}
%\hfill
%\subfigure[Title A]{\includegraphics[width=0.49\linewidth]{fig00.pdf}}
%\hfill
%\subfigure[Title B]{\includegraphics[width=0.49\linewidth]{fig 1_2.pdf}}
%\hfill
%\caption{Title for both}
%\end{figure}

\begin{figure}
    \centering
    \includegraphics[width=0.6\linewidth]{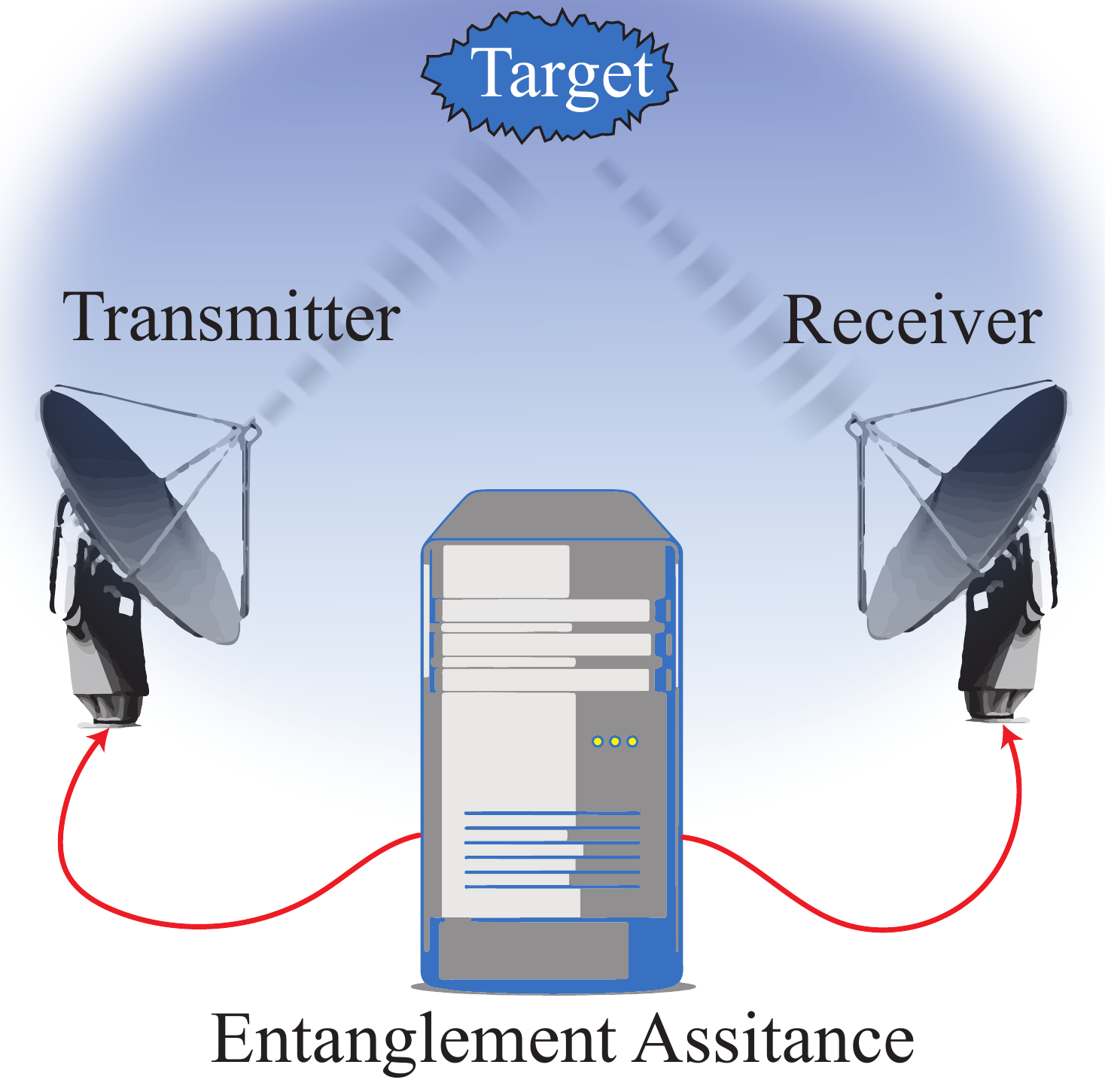}
    \caption{Concept of entanglement-assisted target detection. Target surface can be rough causing fading effects.}
    \label{sch1}
\end{figure}
\begin{figure}[t]
    \centering
    \includegraphics[width=0.95\linewidth]{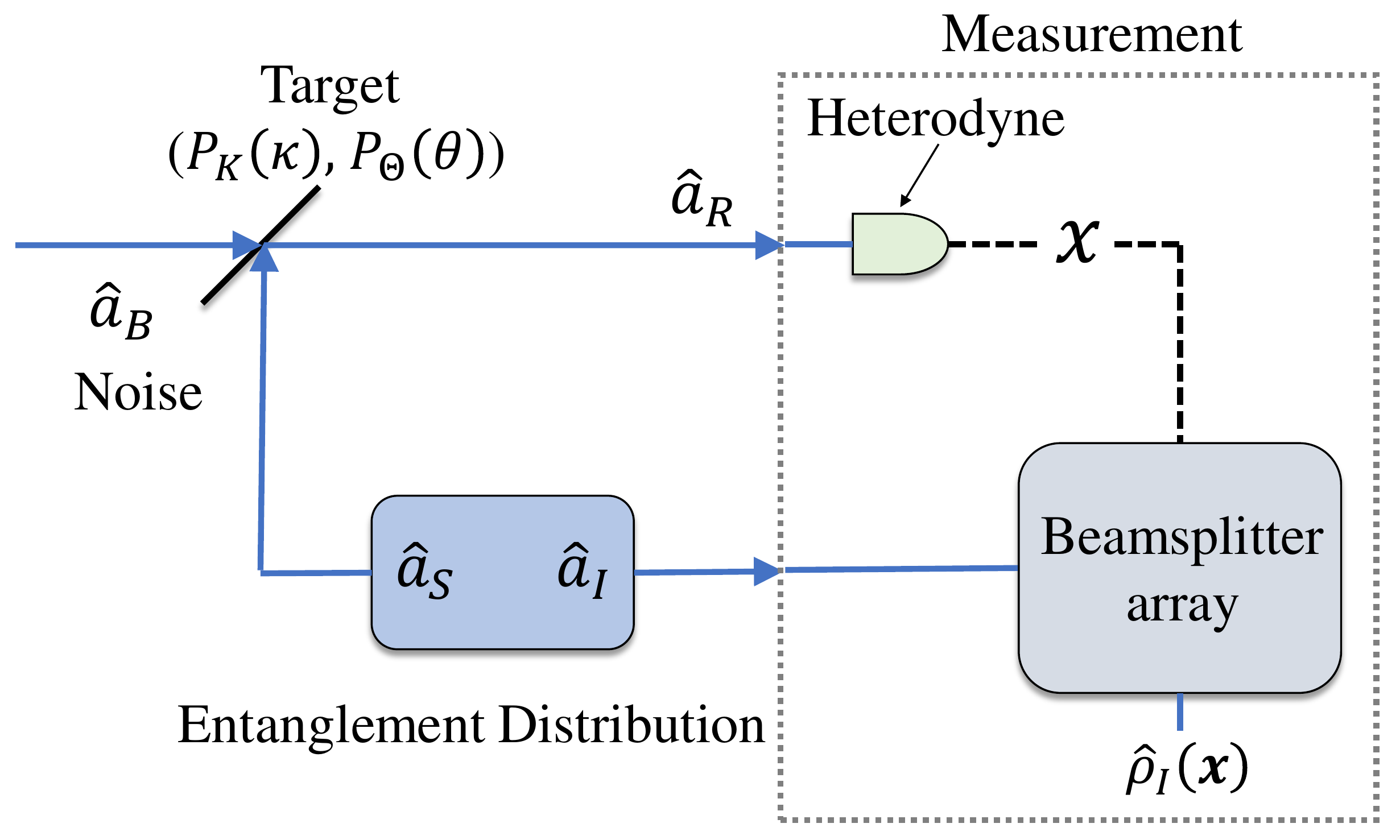}
    \caption{Schematic illustration of C$\veryshortrightarrow$D conversion module.}
    \label{Sch2}
\end{figure}

As shown in Fig.~\ref{sch1}, in an entanglement-assisted QI target detection scenario, the probe signal is entangled with an ancilla. The signal is reflected by a stationary target in a highly lossy and noisy environment before being detected. A properly structured receiver is required to measure the received signal and the ancilla to boost the sensing precision over CI. In the ideal case of a known phase and a fixed target reflectivity, this process can be modeled as an overall phase-shift thermal-loss channel $\Phi_{\kappa, \theta}$~\cite{weedbrook2012gaussian}, with $\kappa$ being the transmissivity and $\theta$ being the phase shift (as shown in Fig.~\ref{Sch2}). 
For an input mode described by the annihilation operators $\hat{a}_{\rm S}$, the received mode is
\be 
\hat{a}_{\rm R}=e^{i\theta}\sqrt{\kappa}\hat{a}_{\rm S}+\sqrt{1-\kappa}\hat{a}_{\rm B},
\label{input_output_main}
\ee 
where the mode $\hat{a}_{\rm B}$ is in a thermal state with mean photon number $N_{\rm E}$ to model the noise.

To model a realistic setting, we consider a target with a time-independent $P_K(\cdot)$-distributed random reflectivity and $P_\Theta(\cdot)$-distributed random phase shift. This leads to the overall quantum channel
\be 
\bar{\Phi}=\int {\rm d\theta}{\rm d\kappa} P_\Theta(\theta)P_K(\kappa) \Phi_{\kappa,\theta}.
\ee 
The target detection hypothesis testing problem is therefore a quantum channel discrimination problem between the channel $\bar{\Phi}$ (fading target present) and a pure noise channel $\Phi_{0,0}$.

To benefit from entanglement in QI, we consider $M$ signal-idler pairs $\{\hat{a}_{S_m},\hat{a}_{I_m}\}_{m=1}^M$, where each pair is in a two-mode squeezed-vacuum (TMSV) state with the wave-function
\be 
\hat{\phi}_{S_m I_m}=\sum_{n=0}^\infty \sqrt{\frac{N_{{\rm S}}^{n}}{(N_{{\rm S}}+1)^{n+1}}} \ket{n}_{S_m}\ket{n}_{I_m}.
\label{eq:state_TMSV}
\ee 
Here $\ket{n}$ is the number state and $N_{{\rm S}}$ is the mean photon number of the signal (or idler) mode.

When the target is present, after the channel $\bar{\Phi}$, the density operator of the return and idler field is
\be
\hat{\rho}_{\rm RI}=\int {\rm d\theta}{\rm d\kappa} P_\Theta(\theta)P_K(\kappa)\hat{\rho}_{\rm RI}(\theta,\kappa).    
\ee
Here the state $\hat{\rho}_{\rm RI}(\theta,\kappa)$ describes the $M$ return-idler pairs $\{\hat{a}_{R_m},\hat{a}_{I_m}\}_{m=1}^M$ from channel $\Phi_{\kappa,\theta}$, each maintaining a phase-sensitive cross-correlation 
$ 
\expval{\hat{a}_{R_m} \hat{a}_{I_m}}=e^{i \theta} C_p
$
with the amplitude $C_p\equiv \sqrt{\kappa    N_{{\rm S}} \left(  N_{{\rm S}}+1\right)}$.

\section{Analyses of correlation-to-displacement conversion module
}

As shown in Fig.~\ref{Sch2}, in a ${\rm C}\veryshortrightarrow {\rm D}$ conversion module~\cite{shi2022}, we perform heterodyne measurement on each return mode and retain the idlers for further information processing. In general, the measurement can be described by positive operator-valued measure (POVM) elements {$\hat{E}^\dagger_{{\bm x}}\hat{E}_{{\bm x}}$} satisfying the completeness relation $\int {d}^{2M}{{\bm x}} \,\hat{E}^\dagger_{{\bm x}}\hat{E}_{{\bm x}} =\hat{I}$, where the overall measurement result across the $M$ returns ${\bm x}=(x_1, \cdots, x_M)^T $ with each $x_m$ being complex.

The corresponding probability of having measurement result $\bm X=\bm x$ is given by
\begin{align}
P_{\bm X}(\bm x)&={\rm Tr}(\hat{\rho}_{\rm RI}\hat{E}^\dagger_{\bm x}\hat{E}_{\bm x} )
\\
&=\int {\rm d\theta}{\rm d\kappa} P_\Theta(\theta)P_K(\kappa)P_{\bm X|\Theta,K}(\bm x|\theta,\kappa),
\end{align} 
with $P_{{\bm X}|\Theta,K}({\bm x}|\theta,\kappa)={\rm Tr}(\hat{\rho}_{\rm RI}(\theta,\kappa)\hat{E}^\dagger_{\bm x}\hat{E}_{\bm x} )$ as the conditional probability when the channel is $\Phi_{\kappa,\theta}$. For a given fixed phase and reflectivity, the distribution has been solved in Ref.~\cite{shi2022} as a complex Gaussian distribution with variance $2\sigma_\kappa^2=\kappa N_{{\rm S}}+(1-\kappa)N_{\rm E}+1$, i.e.,
\be 
P_{{\bm X}|\Theta,K}({\bm x}|\theta,\kappa) =  
g(|\bm x|,\sigma_k),
\label{eq:p_M}
\ee 
where we define $g(x,\sigma)= e^{-x^2/2\sigma^2}/(2\pi\sigma^2)^M$.
Note that $P_{{\bm X}|\Theta,K}({\bm x}|\theta,\kappa)$ does not depend on the phase shift $\theta$; therefore, we obtain the unconditional distribution of the measurement result as
\be
P_{\bm X}({\bm x})=\int {\rm d\kappa} P_{K,{\bm X}}(\kappa,{\bm x}),
\label{PM_def}
\ee
with $P_{K,{\bm X}}(\kappa,{\bm x})\equiv P_K(\kappa)P_{{\bm X}|\Theta,K}({\bm x}|\theta,\kappa)=P_K(\kappa) g(|\bm x|,\sigma_k)$. At the same time, the conditional distribution can be obtained as
\begin{align}
P_{K|\bm X}(\kappa|\bm x)=\frac{P_K(\kappa) g(|\bm x|,\sigma_k)}{\int {\rm d}\kappa P_K(\kappa) g(|\bm x|,\sigma_k)}\equiv f(\kappa,|\bm x|),
\label{f_func}
\end{align}
which is only a function of the module $|\bm x|$ and $\kappa$.

%$P_{\bm X}(\bm x)$ and $P_{{\bm X}|\Theta,K}({\bm x}|\theta,\kappa)$ are the probability and conditional probability that the return mode is projected to state $\ket{{\bm X}}=\otimes_m\ket{{\bm X}_m}$, respectively.

Conditioned on the measurement result of the return mode, the signal-idler joint state is projected to
\begin{align}
&\hat{\rho}_{\rm RI}'({\bm x})=\frac{\hat{E}_{\bm x}\hat{\rho}_{\rm RI}\hat{E}^\dagger_{\bm x}}{P_{\bm X}({\bm x})}
% \\
% &=\frac{\int {\rm d\theta}{\rm d\kappa} P_\Theta(\theta)P_K(\kappa)\hat{E}_{\bm x} \hat{\rho}_{\rm RI}(\theta,
% \kappa)\hat{E}^\dagger_{\bm x} }{P_{\bm X}(\bm x)}
\\
&=\int {\rm d\theta}{\rm d\kappa}\frac{ P_\Theta(\theta)P_{K,{\bm X}}(\kappa,{\bm x})}{P_{\bm X}(\bm x)} \hat{\rho}_{\rm RI}(\theta,\kappa|{\bm x}),
\end{align}
where the conditional state
\be 
\hat{\rho}_{\rm RI}(\theta,\kappa|{\bm x})=\frac{\hat{E}_{\bm x} \hat{\rho}_{\rm RI}(\theta,\kappa)\hat{E}^\dagger_{\bm x} }{P_{{\bm X}|\Theta,K}({\bm x}|\theta,\kappa)}
\ee 
is identical to the return state after the heterodyne detection, when the target has a fixed phase shift $\theta$ and a reflectivity $\kappa$~\cite{shi2022}. Therefore, the idler modes of $\hat{\rho}_{\rm RI}(\theta,\kappa|{\bm x})$ is in product of displaced thermal state 
\be 
{\rm Tr}_R[\hat{\rho}_{\rm RI}(\theta,\kappa|{\bm x})]=\otimes_m \hat{\rho}_{d_m, E_\kappa}.
\ee 
The complex displacement of idler conditioned on the measurement result is
$
d_m=\mu_\kappa{\rm e}^{{\rm i}\theta}{\bm x}_m^{*},
$
with 
\be 
\mu_\kappa=\frac{\sqrt{\kappa N_{{\rm S}}(N_{{\rm S}}+1)}}{[\kappa N_{{\rm S}}+(1-\kappa)N_{\rm E}+1]},
\ee 
and the thermal noise mean photon number 
\be 
E_\kappa=\frac{(1-\kappa)(1+N_{\rm E})N_{{\rm S}}}{[\kappa N_{{\rm S}}+(1-\kappa)N_{\rm E}+1]}.
\ee 
Conditioned on phase $\theta$ and reflectivity $\kappa$, one can apply the beamsplitter array strategy in Ref.~\cite{shi2022} on the idler modes with the weights of the beamsplitter properly chosen based on the heterodyne detection result (indepedent of $\theta$ or $\kappa$), producing a one-mode displaced thermal state with the complex displacement,
$
d=\sum \omega_m d_m=\mu_\kappa{\rm e}^{{\rm i}\theta}|{\bm x}|,
$
where the weight $\omega_m={\bm x}_m/|{\bm x}|$ is independent of $\kappa,\theta$. The mean photon number of the displaced thermal state is still $E_\kappa$. Considering the phase shift and reflectivity distribution, the unconditional output state of the single output mode is
\begin{equation}
\hat{\rho}_{\rm I}({\bm x})=\int {\rm d\kappa} P_{K|{\bm X}}(\kappa|{\bm x})
\hat{\rho}_{\rm I,\kappa}({\bm x})
. 
\label{rho_I}
\end{equation}
where the conditional state 
\be 
\hat{\rho}_{\rm I,\kappa}({\bm x})\equiv \int {\rm d\theta} \,P_\Theta(\theta)\hat{\rho}_{\mu_\kappa{\rm e}^{{\rm i}\theta}|{\bm x}|, E_\kappa}. 
\label{rho_I_k}
\ee 
Note that when the phase is uniform random in $[0,2\pi)$, $\hat{\rho}_{\rm I,\kappa}({\bm x})$ is photon-number diagonal (see Appendix~\ref{diagonal}).
Similar to Eq.~(3) of Ref.~\cite{shi2022}, the error probability performance limit of QI based on the ${\rm C}\veryshortrightarrow {\rm D}$ conversion module is therefore
\begin{align}
P_{\rm C\veryshortrightarrow D}&=\int {\rm d}^{2M}{\bm x} P_{\bm X}(\bm x)P_{\rm H}\left[\hat{\rho}_{0,N_{{\rm S}}},\hat{\rho}_{\rm I}\left({\bm x}\right)\right].
\label{HEL_general}
\end{align}

Noticing that the state $\hat{\rho}_{\rm I}({\bm x})$ and the distribution $P_{\bm X}(\bm x)$ are only functions of the amplitude $|\bm x|$ and making use of Eqs.~\eqref{f_func} and~\eqref{eq:p_M} explicitly, we can further simplify the result via integrating out $2M-1$ degree of freedom to obtain
\begin{align}
P_{\rm C\veryshortrightarrow D}
=
\int {\rm d}x P_X(x)
P_{\rm H}\left[\hat{\rho}_{0,N_{{\rm S}}},\hat{\rho}_{\rm I}\left(x\right)\right]
\label{HEL_general_simple}
\end{align}
Here
\begin{align}
P_X(x)=\frac{2\pi^M}{\Gamma(M)}\int {\rm d}\kappa P_{K}(\kappa) x^{2M-1}\, g(x,\sigma_k).
\label{Px}
\end{align}
is the distribution of the module of measurement result $\bm x$, and the corresponding conditional state
\begin{equation}
\hat{\rho}_{\rm I}(x)=\int {\rm d\theta}{\rm d\kappa} 
P_\Theta(\theta)f(\kappa,x)
\hat{\rho}_{\mu_\kappa{\rm e}^{{\rm i}\theta}x, E_\kappa}. 
\label{rho_I_x}
\end{equation}

% where $A=\int {\rm d}x P(x)x^{2M-1}=2\pi^M/\Gamma(M)$, $\Gamma$ is the gamma function. In the above derivation we have set $x=|{\bm x}|$ and used the properties that $P_{\bm X}(\bm x)$ and $\hat{\rho}_{\rm I}({\bm x})$ only depend on the amplitudes, and therefore we defined $P(|{\bm x}|)\equiv P_{\bm X}(\bm x)$ and $\hat{\rho}_{\rm I}(|{\bm x}|)\equiv \hat{\rho}_{\rm I}({\bm x})$. 

\section{Performance for random phase model (known reflectivity)}

\subsection{Evaluating the performance of conversion module}

To understand the effect of phase noise, we begin with the scenario of uniformly distributed phase shift and a fixed known reflectivity $\kappa$. Therefore, the phase noise distribution $P_\Theta(\theta)=1/2\pi$ and the reflectivity is a delta-function, $P_K(\kappa')=\delta(\kappa'-\kappa)$. Consequently, $\hat{\rho}_{\rm I}=\hat{\rho}_{\rm I,\kappa}$ in Eq.~\eqref{rho_I} is diagonal in the number basis 
% \begin{equation}
% \hat{\rho}_{\rm I,\kappa}(x)=\int {\rm d\theta}\frac{1}{2\pi}\hat{\rho}_{\mu_\kappa{\rm e}^{{\rm i}\theta}x, E_\kappa}. 
% \label{rho_IK}
% \end{equation}
regardless of the target's presence or absence. Therefore, photon counting is the optimal measurement and the error probability performance limit can be analytically solved from Eq.~\eqref{HEL_general_simple} and Eq.~\eqref{Px}, 
\be 
P_{\rm C\veryshortrightarrow D}
=\int{\rm d}y_{\kappa}P_{\chi^2}^{(2M)}(y_{\kappa})P_{\rm H}\left[\hat{\rho}_{0,N_{{\rm S}}},\hat{\rho}_{{\rm I},\kappa}\left(\sigma_\kappa\sqrt{y_{\kappa}}\right)\right]
\label{HEL}
\ee 
where $P_{\chi^2}^{(2M)}(\cdot)$ is the $\chi^2$ distribution of $2M$ degrees of freedom and we have changed the variable $x$ to $y_{\kappa}= x^2/\sigma_\kappa^2$ from Eq.~\eqref{HEL_general_simple}. At the same time, we can explicitly solve
\be
P_{\rm H}\left[\hat{\rho}_{0,N_{{\rm S}}},\hat{\rho}_{{\rm I},\kappa}\left(\sigma_\kappa\sqrt{y_{\kappa}}\right)\right]=\left[1-\sum_{n: \gamma_{n,\kappa}\left(y_k\right)>0}\gamma_{n,\kappa}\left(y_k\right)\right]/2,
\ee 
where we have defined (see Appendix~\ref{diagonal})
\begin{align}
\gamma_{n,\kappa}(y_{\kappa})&=\frac{N_{{\rm S}}^n}{(1+N_{{\rm S}})^{n+1}}
\nonumber
\\
&-\frac{E^n}{(1+E)^{1+n}}{\rm e}^{-\xi_\kappa y_{\kappa}/E} {_1\Tilde{F}_1\Big[n+1,1,\frac{\xi_\kappa y_{\kappa}}{E(1+E)}\Big]},
\label{gamma_y}
\end{align}
and the summation includes all positive values of $\gamma_n\left(y\right)$.
Here $_1\Tilde{F}_1$ is the regularized confluent hypergeometric
function \cite{shi2020practical} and 
\be 
\xi_\kappa=\mu_\kappa^2\sigma_\kappa^2=\frac{\kappa N_{{\rm S}}(N_{{\rm S}}+1)}{2[\kappa N_{{\rm S}}+(1-\kappa)N_{\rm E}+1]}.
\label{xi_k}
\ee 

Moreover, due to $M\gg1$, the $\chi^2$ distribution in Eq.~\eqref{HEL} can be approximated as a delta function, and we arrive at the analytical result
\begin{align}
P_{\rm C\veryshortrightarrow D}&\approx P_{\rm H}\left[\hat{\rho}_{0,N_{{\rm S}}},\hat{\rho}_{{\rm I},\kappa}\left(\sigma_\kappa\sqrt{2M}\right)\right]
\label{conHel_PH}
\\
&=\left[1-\sum_{n: \gamma_{n,\kappa}\left(2M\right)>0}\gamma_{n,\kappa}\left(2M\right)\right]/2,
\label{conHel}
\end{align}
We numerically verified that the above expression agrees with the exact result with negligible error in all the parameter regions relevant to this paper (see Appendix \ref{largeM}).

\begin{figure}
    \centering
    \includegraphics[width=0.8\linewidth]{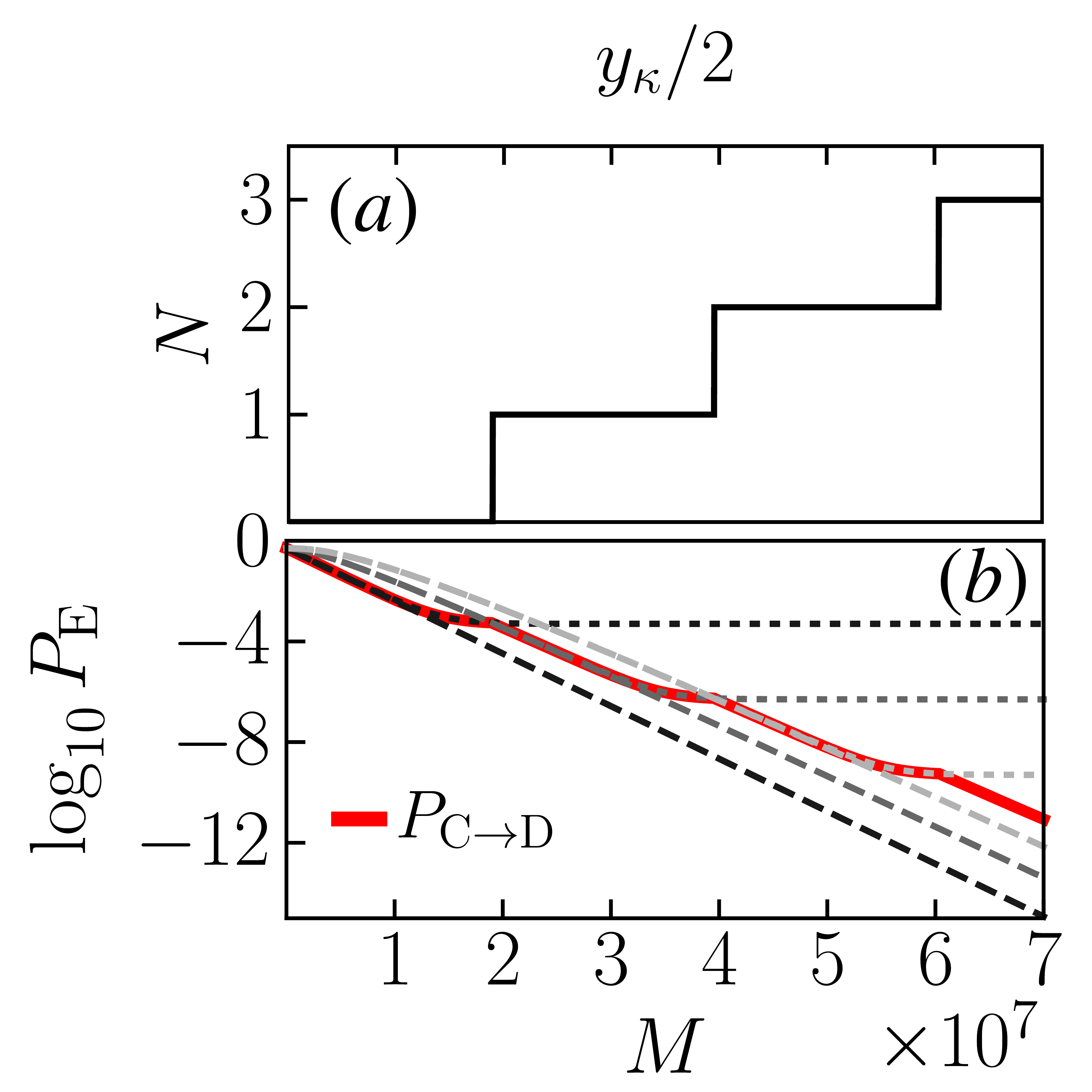}
    \caption{(a) Optimal decision threshold of the photon counts $N$. (b) Error probability
versus the number of copies $M$ with $N_{{\rm S}}=0.001$, $N_{\rm E}=20$ and $\kappa=0.01$. The abrupt changes of $P_{\rm C\veryshortrightarrow D}$ happen when the optimal decision threshold changes in (a). The dashed lines from dark to light represent the error probabilities corresponding to the decision threshold $N=0, 1, 2$, respectively, according to the approximation in Eq.~(\ref{app1}). The dotted lines from dark to light indicate the exact error probabilities for the same thresholds calculated with Eq.~(\ref{app2}).
}
    \label{fig1}
\end{figure}

In Fig.~\ref{fig1} (b), we plot QI performance $P_{\rm C\veryshortrightarrow D}$ as the red curve for the same parameter choice of Refs.~\cite{shi2022,Zhuang2017}. We see abrupt changes in the error probability when the number of modes $M$ increases, due to the integer summation in Eq.~\eqref{conHel}. To better understand the performance, we consider a threshold decision strategy, where one compares the measured photon number against a threshold $N$: target presence is declared if and only if the photon number is larger than $N$. From Eq.~\eqref{conHel}, the error probability of such a threshold decision is 
\be
P_{{\rm C\veryshortrightarrow D},\kappa}^{N}=\frac{1}{2}\left[1-\sum_{n=0}^{N}\gamma_{n,\kappa}\left(2M\right)\right].
\label{app2}
\ee
We plot $P_{{\rm C\veryshortrightarrow D},\kappa}^{N}$ as the dotted lines for different values of $N$ and they agree with $P_{\rm C\veryshortrightarrow D}$ within each continuous sector (solid red curve). The abrupt changes of $P_{\rm C\veryshortrightarrow D}$ also corresponds well with the change in the optimal decision threshold $\argmin_N{P_{{\rm C\veryshortrightarrow D},\kappa}^{N}}$ in Fig.~\ref{fig1} (a).

\begin{figure*}
    \centering
    \includegraphics[width=\linewidth]{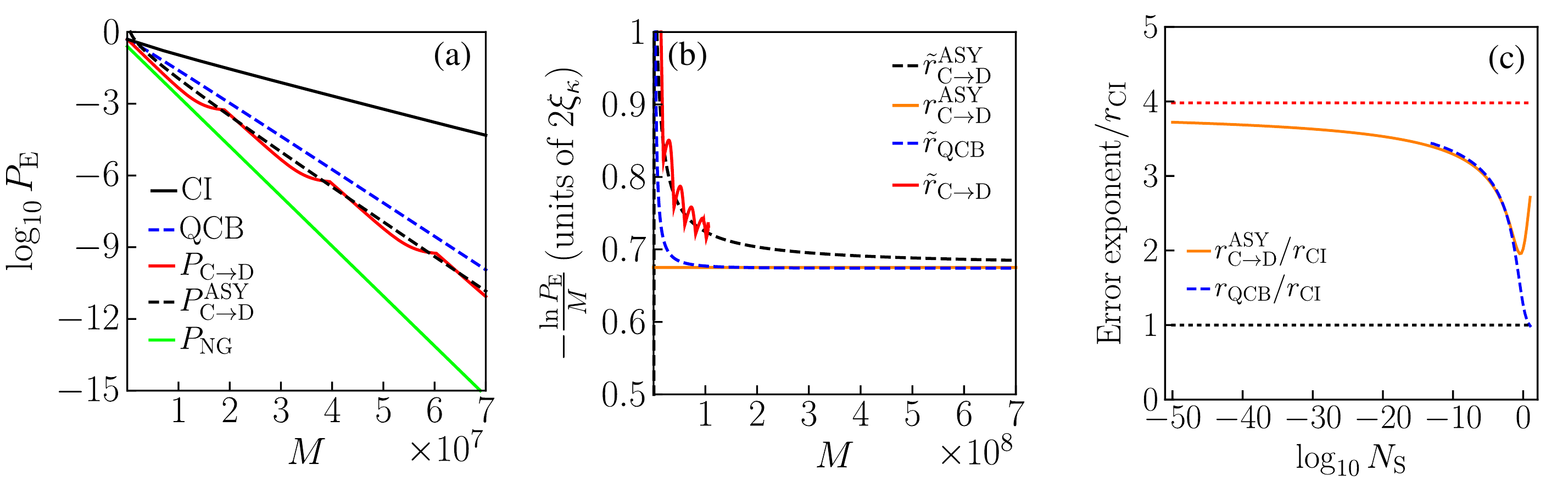}
   \caption{The error performance for the uniform phase and known reflectivity model with the parameters: $N_{{\rm S}}=0.001$, $N_{\rm E}=20$ and $\kappa=0.01$. Note: some lines are only partially plotted in a range of x-axis limited by numerical precision. (a) The error probabilities and bounds thereof as a function of $M$. The red line indicates the error probability limit of the C$\veryshortrightarrow$D conversion. The black dashed line is the asymptote of this probability. The blue and green lines are the QCB (upper bound) and NG lower bound, respectively. The black line indicates the optimum CI’s error probability.
    (b)The asymptotic behaviors of $-\ln{P_{\rm E}}/M$ (whose asymptotic limits are the error exponents) with $P_{\rm E}$ being the error performance of the C$\veryshortrightarrow$D conversion module (red), the asymptote (black, dashed) and the QCB (blue, dashed), respectively. The orange line indicates the error exponent of the asymptote. (c) The error exponent (normalised by the error exponent of the CI) of the asymptote and QCB. When $N_{\rm S}\to 1^-$, the error exponent of asymptote deviates from QCB for the low brightness condition doesn't hold anymore. The red dotted line indicates the 6 dB advantage over CI.}
    \label{fig2}
\end{figure*}

After understanding the performance enabled by the conversion module, now we compare the QI error probability $P_{\rm C\veryshortrightarrow D}$ of Eq.~\eqref{conHel} with that of CI to show the entanglement's advantage. In CI with coherent-state probes, due to the uniform random phase noise, the received state is photon-number diagonal, and the Helstrom limit can be efficiently evaluated (see Appendix~\ref{CI}).
As Fig.~\ref{fig1} already has too many lines, we re-print $P_{\rm C\veryshortrightarrow D}$ (red solid) in Fig.~\ref{fig2}(a) in comparison with the error probability of CI (black solid) and, showing orders of magnitude advantage. In particular, the curves indicate that QI and CI still have different error exponents despite the fully random phase noise, as we will confirm in the next section with asymptotic analyses.

\subsection{Asymptotic results and error exponent}

To better understand the QI performances, and in particular to understand the error exponent in presence of the random phase noise,  we explore asymptotic solutions of $P_{\rm C\veryshortrightarrow D}$.
Considering Eqs.~\eqref{conHel_PH} and~\eqref{rho_I_k} at the low brightness ($N_{{\rm S}}\ll 1$) and low reflectivity ($\kappa\ll 1$) limit, we can approximate the noisy displaced coherent state in Eq.~\eqref{rho_I_k} as a coherent state and $\hat{\rho}_{0,N_{{\rm S}}}$ as a vacuum state.
Therefore, Eq.~\eqref{gamma_y} can be approximated as
$
\gamma_{n,\kappa}(y_{\kappa})=\delta_{n,0}-e^{-\xi_\kappa y_{\kappa}}(\xi_\kappa y_{\kappa})^{n}/n!,
$
where $\delta_{n,0}$ is the Kronecker delta function.

% The error probability is given by the conditional Bayesian probability \cite{zhuang2017fading},
% \begin{align}
% P_N(y)&=\frac{1}{2}P_{F,N}+\frac{1}{2}P_{M,N}
% \\
% &=\frac{1}{2}\sum_{n=0}^N p_n+\frac{1}{2}\sum_{n=N+1}^\infty p'_n=\frac{1}{2}\sum_{n=0}^N p_n,
% \label{Nep}
% \end{align}
% \QZ{**can the above equation directly follow from Eq.~\eqref{conHel}?}
% with the assumption of equally-likely target absence or presence. Here $P_{F,N}$ is the false-alarm probability which describes the chance that an absent target is declared present; $P_{M,N}$ is the missing probability of deciding target is present when it is actually absent. $p'_n$ is the diagonal term of the density matrix of the vacuum, which has zero value with the exception of $p'_0$. Therefore, $P_{M,N}$ is zero for any decision threshold. 
Given a threshold $N$, from Eq.~\eqref{app2}, the error probability of the C$\veryshortrightarrow$D conversion module in large-$M$ limit is \begin{align}
P_{{\rm C\veryshortrightarrow D},\kappa}^{N}
%=\int{\rm d}y\chi^2_{2M}(y)P_N(y)\approx P_N(2M)
%\\
\approx\frac{1}{2}\sum_{n=0}^N p_n, \mbox{\ with $p_n=e^{-2\xi_{\kappa} M}(2\xi M)^{n}/n!$}
\label{app1}
\end{align} 
and the minimum error probability $P_{\rm C\veryshortrightarrow D}=\min_N P_{{\rm C\veryshortrightarrow D},\kappa}^{N}$. When the photon number threshold $N=0$, it is just the error probability of Kennedy receiver and $P_{{\rm C\veryshortrightarrow D},\kappa}^{0}=(1/2)e^{-2\xi_{\kappa} M}$ ~\cite{Kennedy_1972}. The dashed lines in Fig.~\ref{fig1}(b) show the approximated error probabilities for the decision threshold $N=0,1,2$, respectively. We see a good recovery of the $P_{\rm C\veryshortrightarrow D}$ (solid red curve) in each continuous sector, which allows us to proceed with the asymptotic analyses.

Next, we obtain the asymptotic optimal decision threshold. Consider Eq.~\eqref{conHel_PH}, now we treat $\hat{\rho}_{0,N_{{\rm S}}}$ as thermal state again. Its density matrix is diagonal, with elements $p'_n=N_{{\rm S}}^n/(1+N_{{\rm S}})^{n+1}$. The optimal threshold is determined by solving $p_N=p'_N$, where $p_N$ is defined in Eq.~\eqref{app1}, 
we obtain
\be
N\approx \frac{2\xi_{\kappa} M}{\epsilon},
\label{Ny}
\ee
where $\epsilon=-W_{-1}(-N_{{\rm S}}/e)\gg 1$ and $W_{-1}$ is Lambert $W$ function. The approximation holds when $M\gg 1$. An asymptote of the Helstrom limit $P_{\rm C\veryshortrightarrow D}^{\rm ASY}$ can be obtained by substituting Eq.~(\ref{Ny}) into Eq.~(\ref{app1}) and its error exponent can be obtained as (see  Appendix~\ref{asym})
\be 
r_{\rm C\veryshortrightarrow D}^{\rm ASY}=\lim_{M\to\infty}\tilde{r}_{\rm C\veryshortrightarrow D}^{\rm ASY}(M) =[1-{\rm ln}(e\epsilon)/\epsilon]2\xi_{\kappa},
\label{rCD}
\ee 
where we defined the finite-$M$ exponent
\be
\tilde{r}_{\rm C\veryshortrightarrow D}^{\rm ASY}\equiv -{\rm ln}P_{\rm C\veryshortrightarrow D}^{\rm ASY}(M)/M.
\ee

Now we evaluate $P_{\rm C\veryshortrightarrow D}^{\rm ASY}$ in Fig.~\ref{fig2}(a) as the black dashed curve. Indeed, we see a good agreement with $P_{\rm C\veryshortrightarrow D}$ of Eq.~\eqref{conHel} (red solid). To understand the error exponent, we plot the error probability in a logarithmic version $-\ln P_{\rm E}/M$ in units of $2\xi_\kappa$ [see Eq.~\eqref{xi_k}] versus the number of modes $M$ in Fig.~\ref{fig2}(b). As expected, $\tilde{r}_{\rm C\veryshortrightarrow D}^{\rm ASY}$ (black dashed) approaches $r_{\rm C\veryshortrightarrow D}^{\rm ASY}$ (orange solid) in the large $M$ limit. The exact results $P_{\rm C\veryshortrightarrow D}$ (red solid) agrees well with $\tilde{r}_{\rm C\veryshortrightarrow D}^{\rm ASY}$, however, its evaluation is limited to rather small $M$ due to numerical precision constraints. 

With the error exponent $r_{\rm C\veryshortrightarrow D}^{\rm ASY}$ in hand, we can now compare with the error exponent of CI $r_{\rm CI}=\lim_{M\to\infty}-\ln \left(P_{\rm CI}\right)/M$ (see Appendix~\ref{CI} for the calculation of $P_{\rm CI}$) to understand the quantum advantage in the error exponent under different signal brightness. 
As shown in Fig.~\ref{fig2}(c), $r_{\rm C\veryshortrightarrow D}^{\rm ASY}$ (orange solid) is always larger than $r_{\rm CI}$, confirming quantum advantage, moreover, the error exponent ratio approaches six decibels (indicated by the red dotted line) as $N_S$ approaches zero, although the rate of convergence is very slow. This can be confirmed analytically from Eq.~\eqref{rCD} via 
\be 
\lim_{N_S\to 0} r_{\rm C\veryshortrightarrow D}^{\rm ASY}=2\xi_{\kappa} \simeq \kappa N_S/N_E.
\ee 
As we have $r_{\rm CI}\lesssim \kappa N_S/4N_E$, there is indeed a six-decibel advantage of QI over CI. From the numerical results as well as asymptotic analyses, we see that in the weak signal limit, phase noise essentially does not change the error exponent, compared to the case without phase noise~\cite{tan2008quantum,shi2022}.

\subsection{Upper and lower bounds}

Finally, we provide additional comparison of the QI performance with upper and lower bounds. We obtain upper bound from the asymptotically tight QCB~\cite{Audenaert2007,Pirandola2008} and lower bound from the Nair-Gu (NG) bound~\cite{nair2020fundamental}.

Given any two quantum state $\hat{\rho}_0,\hat{\rho}_1$, the QCB $P_{\rm QCB}(\hat{\rho}_0,\hat{\rho}_1)=(1/2){\rm inf}_{s\in[0,1]}Q_s$, where $Q_s={\rm Tr}(\hat{\rho}_0^s\hat{\rho}_1^{1-s})$, is an asymptotically tight upper bound for the Helstrom limit $P_{\rm H}\left[\hat{\rho}_0,\hat{\rho}_1\right]$.
Therefore, for the uniform phase and known reflectivity model, we can apply the QCB on the Helstrom limit $P_{\rm H}\left[\hat{\rho}_{0,N_{{\rm S}}},\hat{\rho}_{{\rm I},\kappa}\left(\sigma_\kappa\sqrt{y_{\kappa}}\right)\right]$ in Eq.~\eqref{conHel_PH} to obtain the upper bound
\begin{align}
&P_{\rm C\veryshortrightarrow D}\le P_{\rm QCB,U}
% \\
% &\equiv \int{\rm d}yP_{\chi^2}^{(2M)}(y_{\kappa})\frac{{\rm inf}_{s\in[0,1]}{\rm Tr}[\hat{\rho}_{0,N_{{\rm S}}}^s\hat{\rho}_{\rm I}^{1-s}(\sigma_\kappa\sqrt{y_{\kappa}})]}{2}\nonumber
\equiv  \frac{{\rm inf}_{s\in[0,1]}{\rm Tr}[\hat{\rho}_{0,N_{{\rm S}}}^s\hat{\rho}_{{\rm I},\kappa}^{1-s}(\sigma_\kappa\sqrt{2M})]}{2}.
\end{align}
Here both $\hat{\rho}_{0,N_{{\rm S}}}$ and $\hat{\rho}_{{\rm I},\kappa}$ are diagonal in the number state basis and therefore can be efficiently evaluated.

Nair and Gu derived a lower bound on the error probability of quantum illumination (QI)~\cite{nair2020fundamental} target detection assisted by arbitrary form of entanglement. As this is the lower bound in the ideal case, it also holds as a lower bound in presence of additional noise. Consider $M$ probes with mean photon number $N_{{\rm S}}$, we then have
\be
P_{\rm C\veryshortrightarrow D} \geq P_{\rm NG}=\frac{1}{4}e^{-\beta M N_{{\rm S}}},
\ee
where $\beta=-\ln[1-\kappa/(N_E(1-\kappa)+1)]$. 

We plot the upper bound $P_{\rm QCB,U}$ (blue dashed) and lower bound $P_{\rm NG}$ (green solid) in Fig.~\ref{fig1} (a). Meanwhile, we also plot the QCB error exponent $r_{\rm QCB}\equiv\lim_{M\to\infty}-\ln P_{\rm QCB,U}/M$ and $\tilde{r}_{\rm QCB}\equiv -\ln P_{\rm QCB,U}/M$ in Fig.~\ref{fig1} (b) and (c). Indeed, we see that QCB verifies our previous asymptotic evaluations. 

\begin{figure}[t]
    \centering
    \includegraphics[width=0.6\linewidth]{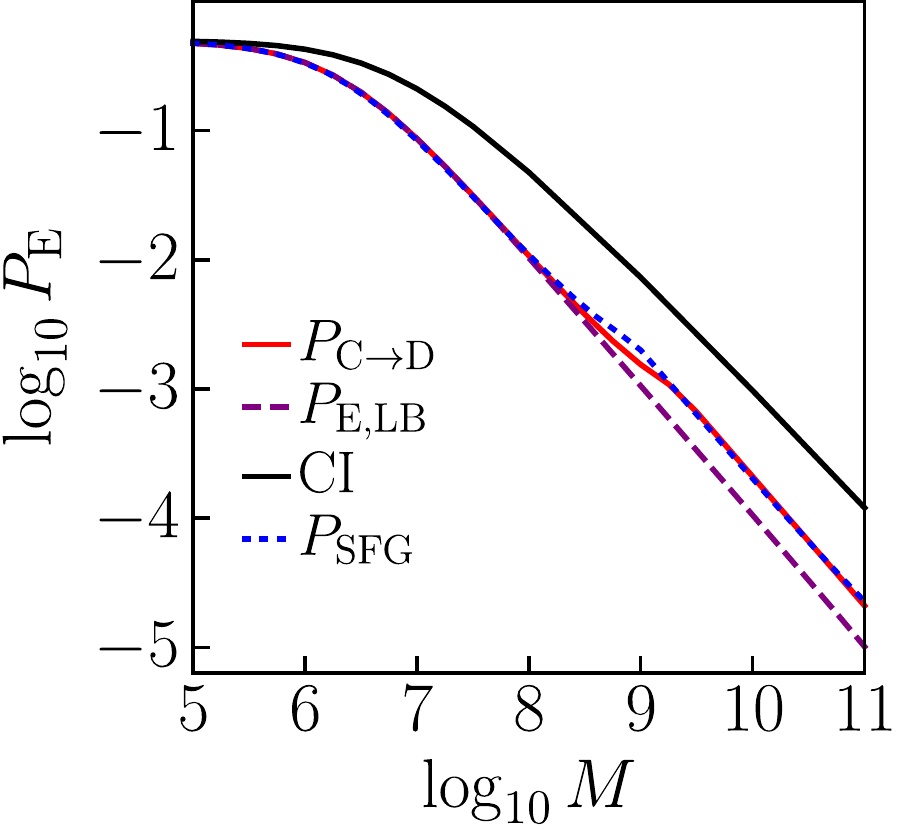}
    \caption{The error performance for the Rayleigh-fading model. The parameters are $N_{{\rm S}}=0.001$, $N_{\rm E}=20$ and $\bar{\kappa}=0.01$.
     Comparison of the achievable error performance given by Eq.~(\ref{heli}) (red), the lower bound given by Eq.~(\ref{LB}) (purple, dashed), the optimum error probability for CI (black) and the error performance of SFG receiver (blue, dotted).
    }
    \label{fig5}
\end{figure}

\section{Performance for Rayleigh-fading Model}

With the performance degradation from phase noise well understood, now we consider Rayleigh-fading targets, where the target has a Rayleigh-distributed reflectivity besides a uniform random phase, i.e.,
\be
P_K(\kappa)= e^{-\kappa/\bar{\kappa}}/\bar{\kappa},
\label{PDFK}
\ee
with $\bar{\kappa}$ being the average reflectivity of the target. Note the above distribution is up to a cut-off so that $\kappa\in[0,1]$.  

As Eqs.~(\ref{rho_I}) and (\ref{HEL_general}) are now difficult to calculate numerically, to understand the QI performance for Rayleigh-fading targets, we consider lower bounds and achievable performance (upper bounds).

\subsection{Lower bound}

%\begin{align}
%P_{\rm C\veryshortrightarrow D}&=\int {\rm d}^{2M}{\bm x} P_{\bm X}(\bm x)P_{\rm H}\left[\hat{\rho}_{0,N_{{\rm S}}},\hat{\rho}_{\rm I}\left({\bm x}\right)\right].
%\\ 
%&\ge 
%\int {\rm d}^{2M}{\bm x} 
%{\rm d\kappa} P_{K,{\bm X}}(\kappa,{\bm x})
%P_{\rm H}\left[\hat{\rho}_{0,N_{{\rm S}}},\hat{\rho}_{\rm I,\kappa}(\bm x)\right]
%\end{align}

Applying concavity of the Helstrom limit (see Lemma 1 in \cite{zhuang2017fading}) to Eqs.~(\ref{rho_I}) and (\ref{HEL_general}), we have
\begin{align}
&P_{\rm C\veryshortrightarrow D}
\geq P_{\rm E,LB}\equiv
\int {\rm d}^{2M}{\bm x} 
{\rm d\kappa} P_{K,{\bm X}}(\kappa,{\bm x})P_{\rm H}\left[\hat{\rho}_{0,N_{{\rm S}}},\hat{\rho}_{\rm I,\kappa}(\bm x)\right]\nonumber
\\
&=\int{\rm d}\kappa {\rm d}y_{\kappa} \frac{1}{\bar{\kappa}}e^{-\kappa/\bar{\kappa}}P_{\chi^2}^{(2M)}(y_{\kappa}) P_{\rm H}\left[\hat{\rho}_{0,N_{{\rm S}}},\hat{\rho}_{{\rm I},\kappa}\left(\sigma_\kappa\sqrt{y_{\kappa}}\right)\right]\nonumber
\\
&\approx\int {\rm d}\kappa  \frac{1}{\bar{\kappa}}e^{-\kappa/\bar{\kappa}} P_{\rm H}\left[\hat{\rho}_{0,N_{{\rm S}}},\hat{\rho}_{{\rm I},\kappa}(\sqrt{2M}\sigma_{\kappa})\right].
\label{LB}
\end{align}
In the last step, we have taken the approximation at the $M\gg 1$ limit, similar to Eq.~\eqref{conHel_PH}. Now Eq.~\eqref{LB} can be evaluated via an approach similar to Eq.~\eqref{HEL}.

\subsection{Achievable performance}
We then explore an achievable performance of the ${\rm C \veryshortrightarrow D}$ conversion module for the Rayleigh-fading model. Upon the heterodyne measurement results on the return $\bm x$, we perform direct photon counting on the idler output in state $\hat{\rho}_{\rm I}({\bm x})$ from the conversion module, then finish with a threshold decision strategy at a fixed threshold independent of $\bm x$. With the decision threshold optimized, the error probability can be expressed as
\begin{align}
&P_{\rm C\veryshortrightarrow D}= P_{\rm H}\bigg[\hat{\rho}_{0,N_{{\rm S}}},\int {\rm d}^{2M}{\bm x}P_{\bm X}(\bm x)\hat{\rho}_{\rm I}\left({\bm x}\right)\bigg]\nonumber
\\
&=P_{\rm H}\bigg[\hat{\rho}_{0,N_{{\rm S}}},  \int{\rm d}\kappa {\rm d}y_{\kappa} \frac{1}{\bar{\kappa}}e^{-\kappa/\bar{\kappa}}P_{\chi^2}^{(2M)}(y_{\kappa})\hat{\rho}_{{\rm I},\kappa}\left(\sigma_\kappa\sqrt{y_{\kappa}}\right)\bigg]
\nonumber
\\
&\approx P_{\rm H}\left[\hat{\rho}_{0,N_{{\rm S}}},\int {\rm d}\kappa \frac{1}{\bar{\kappa}} e^{-\kappa/\bar{\kappa}}\hat{\rho}_{{\rm I},\kappa}(\sqrt{2M}\sigma_{\kappa})\right],
\label{heli}
\end{align}
where in the last step the measurement distribution is approximated as a delta-function at the large $M$ limit. 

Fig.~\ref{fig5} plots the achievable performance $P_{\rm C\veryshortrightarrow D}$ (red solid), the lower bound $P_{\rm E,LB}$ (purple dashed) and the optimum CI’s error probability (black solid, see Appendix~\ref{CI}) versus the number of modes. We see that the quantum advantage over CI persists for the Rayleigh-fading model, although it is further reduced when compared with the random phase model. The plot also shows that our results agree with the QI detection for the Rayleigh-fading targets with the SFG reception $P_{\rm SFG}$~\cite{zhuang2017fading} (blue dashed), where the error probability decays with the number of modes in a polynomial fashion. Indeed, we find the achievable result $P_{\rm C\veryshortrightarrow D}$ of the conversion module agrees fairly well with $P_{\rm SFG}$. While the SFG results require an approximate solution of a complex quantum nonlinear optical process, the conversion module's achievable performance is almost exact, and requires little effort in calculations.

%The error performance of the C$\veryshortrightarrow$D conversion module for Reyleigh-fading target estimated by the achievable performance and the lower bound qualitatively agrees with the SFG receiver (see Ref.~\cite{zhuang2017fading}), as shown in Fig.~\ref{fig8} (b). 

\section{CONCLUSIONS}
We study the entanglement-assisted target detection performance of the recently proposed correlation-to-displacement conversion module, in the more practical scenario of random phase noise and reflectivity fluctuation. The results show, in the scenario of only random phase noise, this module still affords six-decibel error exponent advantage over the optimum classical illumination when the signal brightness is small. While in consideration of the Rayleigh reflection, the advantage is much smaller, although being non-zero.

\begin{acknowledgements}
This project is supported by the NSF CAREER Award CCF-2142882, NSF OIA-2134830 and NSF OIA-2040575. QZ also acknowledges support from Defense Advanced Research Projects Agency (DARPA) under Young Faculty Award (YFA) Grant No. N660012014029, National Science Foundation (NSF) Engineering Research Center for Quantum Networks Grant No. 1941583.
\end{acknowledgements}

\appendix

\section{Proof of diagonal density matrix of $\hat{\rho}_{\rm I}$ under uniform phase rotation}
\label{diagonal}

Phase rotation $\hat{a}\to e^{-i\theta}\hat{a}$ on mode $\hat{a}$ is described by the unitary $\hat{R}(\theta)=\exp\left[-i\theta \hat{a}^\dagger\hat{a}\right]$.
Under a uniform random phase, any single-mode input state becomes number-state diagonal, because 
\begin{align}
\expval{m|\int {\rm d}\theta\, \hat{R}(\theta)\hat{\rho}\hat{R}^\dagger(\theta)|n}&=\int {\rm d}\theta\, e^{-i\theta(m-n) }\expval{m|\hat{\rho}|n}
\nonumber
\\
&\propto \delta_{mn}\expval{n|\hat{\rho}|n},
\end{align}
where we utilized the fact $\hat{R}(\theta)\ket{n}=e^{-in\theta}\ket{n}$.

In the case of displaced thermal state, we have~\cite{shi2020practical}
\begin{align}
&\bra{m}\hat{\rho}_{{\rm I},\kappa}(x)\ket{n}=\int {\rm d\theta}\frac{1}{2\pi}\bra{m}\hat{\rho}_{\mu_\kappa x{\rm e}^{{\rm i}\theta}, E_\kappa}\ket{n}\nonumber
\\
&=\frac{\delta_{m,n}E^n}{(1+E)^{1+n}}{\rm e}^{-|\mu_\kappa x|^2/E} {_1\Tilde{F}_1\Big[n+1,1,\frac{|\mu_\kappa x|^2}{E(1+E)}\Big]}.   
\end{align}

\section{Optimum performance limit of classical illumination }
\label{CI}
\begin{figure}
    \centering
    \includegraphics[width=0.6\linewidth]{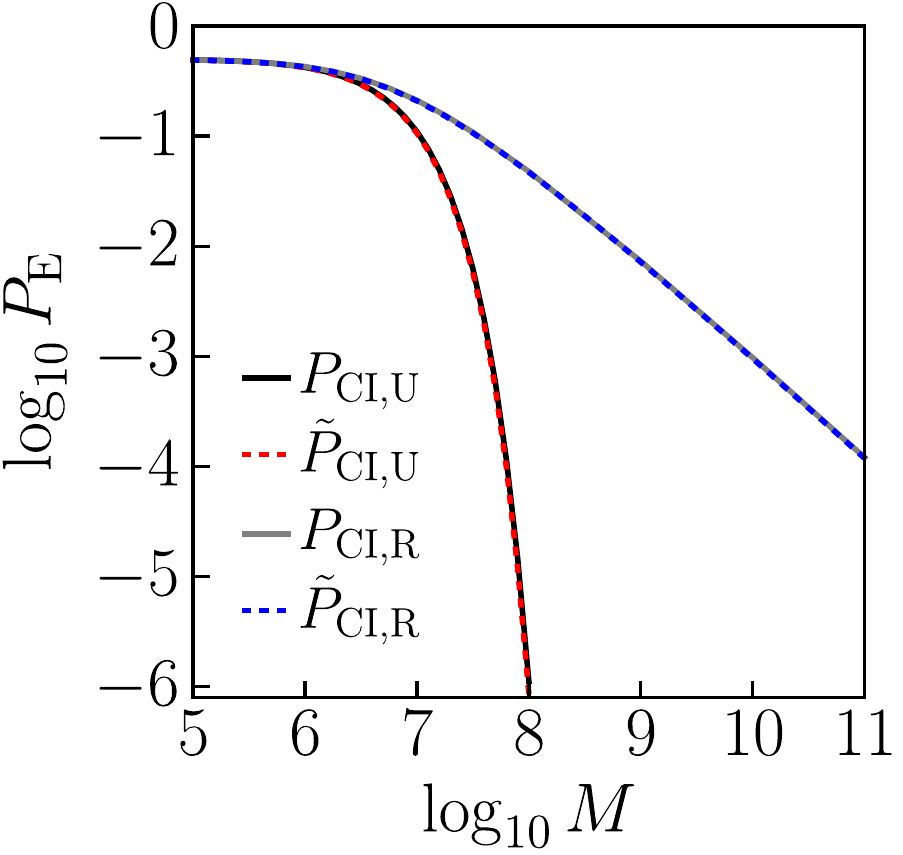}
    \caption{Comparison of the error performances for CI evaluated by Eq.~(\ref{PCI}) (solid) and Eq.~(\ref{PROC}) (dotted), respectively. $P_{\rm CI,U}$ and $\tilde{P}_{\rm CI,U}$ are for the random phase model. $P_{\rm CI,R}$ and $\tilde{P}_{\rm CI,R}$ are for the Rayleigh-fading model.}
    \label{figCI}
\end{figure}
For comparison, the Helstrom limit of classical illumination (CI) is calculated with a coherent-state transmitter. If $\kappa$ and $\theta$ are fixed, the returned mode is in a displaced thermal state $\hat{\rho}_{\sqrt{\kappa N_S},(1-\kappa) N_E}$. When $\kappa$ and $\theta$ are random variables, the output state is then $\int {\rm d\theta}{\rm d\kappa} P_\Theta(\theta)P_K(\kappa)\hat{\rho}_{\sqrt{\kappa N_S},(1-\kappa) N_E}$. Therefore, the performance limit is
\begin{align}
P_{\rm CI}=\frac{1}{2} \left(1-\sum_{n:\gamma_{n,{\rm CI}}>0}\gamma_{n,{\rm CI}}\right),
\label{PCI}
\end{align}
where the summation $\sum$ includes all the positive values of
\begin{align}
\gamma_{n,{\rm CI}}&=\frac{N_E^n}{(1+N_E)^{n+1}}-\int {\rm d}\kappa P_K(\kappa)\frac{E^{\prime n}}{(1+E')^{1+n}}\nonumber\\&\times{\rm e}^{-M\kappa N_{{\rm S}}/E'} {_1\Tilde{F}_1\Big[n+1,1,\frac{M\kappa N_{{\rm S}}}{E'(1+E')}\Big]}.
\end{align}
Here $E'=(1-\kappa) N_E$.

To double check the result, we calculate the performance limit with another method~\cite{zhuang2017fading}: \begin{align}
P_{\rm CI}={\rm min}_{P_{\rm F}^{\rm CI}}\left[P_{\rm F}^{\rm CI}/2+(1-P_{\rm D}^{\rm CI})/2\right]
\label{PROC}
\end{align}
and compare the results. Here the conditional false-alarm probability $P_{\rm F}^{\rm CI}$ denotes the chance that target present is declared when no target is present, and the conditional detection probability $P_{\rm D}^{\rm CI}$ denotes the chance that target present is declared when
a target is present. The relation between $P_{\rm F}^{\rm CI}$ and $P_{\rm D}^{\rm CI}$ is referred to as the receiver operating characteristic (ROC). The ROC for the CI detection of the uniform-phase and known-reflectivity targets is $P_{\rm D}^{\rm CI}=Q(\sqrt{2\kappa M N_S/E'},\sqrt{-2{\rm ln}P_{\rm F}^{\rm CI}})$, where $Q(a,b)$ is the Marcum’s $Q$ function; The CI ROC for the Rayleigh-fading targets is $P_{\rm D}^{\rm CI}=(P_{\rm F}^{\rm CI})^{1/(1+M\bar{\kappa}N_{\rm S}/E')}$~\cite{van2001detection3}. Fig.~\ref{figCI} shows the results calculated with the two methods are consistent.

\section{Large-$M$ approximation}
\label{largeM}

% Since $P_{\chi^2}^{(2M)}(y_{\kappa})\approx (1/\sqrt{2\pi M}){\rm e}^{-\frac{1}{2}\frac{(y_{\kappa}-2M)^2}{4M}}\approx \delta(y_{\kappa}-2M)$ when $M\gg 1$, Eq.~(\ref{conHel}) can be approximated as,
\begin{figure}
    \centering
    \includegraphics[width=0.6\linewidth]{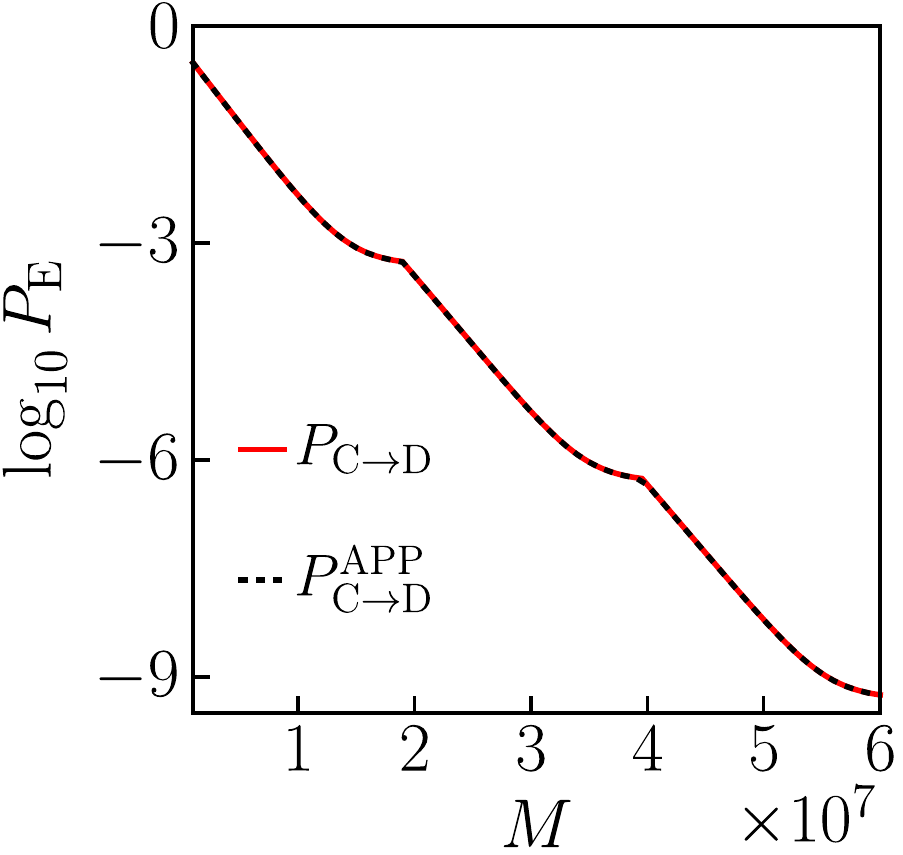}
    \caption{Comparison of the exact (red) and approximated (black, dotted) performance limit for the random phase model in the large-$M$ limit.}
    \label{app}
\end{figure}
%\begin{align}
% P_{\rm C\veryshortrightarrow D}&=\int{\rm d}yP_{\chi^2}^{(2M)}(y_{\kappa})P_H\left[\hat{\rho}_{0,N_{{\rm S}}},\hat{\rho}_{\rm I}\left(\sigma_\kappa\sqrt{y_{\kappa}}\right)\right]\nonumber
% \\
% &=\int{\rm d}yP_{\chi^2}^{(2M)}(y_{\kappa})\left[1-\sum_{n}\gamma'_n\left(y_{\kappa}\right)\right]/2\nonumber
% \\&\approx\int{\rm d}y_{\kappa}\delta(y_{\kappa}-2M)\left[1-\sum_{n}\gamma'_n\left(y_{\kappa}\right)\right]/2\nonumber
% \\&\approx\left[1-\sum_{n}\gamma'_n\left(2M\right)\right]/2.
% \label{exapp}
% \end{align}

Fig.~\ref{app} shows the exact result of Eq.~\eqref{HEL} and approximated performance limits of Eq.~\eqref{conHel} for $M$ in the range of $10^6$ to $6\times10^7$. The maximal deviation for the data we have is $0.25\%$, which happens when $M=10^6$. This approximation is also used in the calculation of the QCB performance for the random phase model and the error performance for the Rayleigh-fading model under the condition of large $M$.

\section{The asymptote of the error probability for random phase model}
\label{asym}
Consider the scenario of uniformly distributed phase shift and a fixed known reflectivity. In the asymptotic limit of low brightness $N_{{\rm S}}\ll 1$ and low reflectivity $\kappa\ll 1$, the optimal decision threshold $N$ is determined by solving
\be
\frac{N_{{\rm S}}^N}{(1+N_{{\rm S}})^{N+1}}=e^{-2\xi_{\kappa} M}(2\xi_{\kappa} M)^{N}/N!,
\ee
which leads to the solution
\be
N\approx-W_{-1}^{-1}\left[-N_{{\rm S}}(\frac{ 1}{1+N_{{\rm S}}})^{1+1/N}\frac{(2\pi N)^{1/N}}{e}\right]2\xi_{\kappa} M.
\label{dec}
\ee
In the derivation above, we have used Stirling's approximation $N!\approx\sqrt{2\pi N}(N/e)^N$. When $N\gg 1$ and $N_S\ll 1$, Eq.~(\ref{Ny}) is obtained by further approximation. Substituting Eq.~(\ref{Ny}) into Eq.~(\ref{app1}), an asymptote of $P_{\rm C\veryshortrightarrow D}$ is obtained,
\begin{align}
P_{\rm C\veryshortrightarrow D}^{\rm ASY}&\approx\frac{1}{2} e^{-2\xi_{\kappa} M}(2\xi_{\kappa} M)^{N}/N!\nonumber\\&\approx \frac{1}{2} e^{-2\xi_{\kappa} M}\frac{(2\xi_{\kappa} M)^{N}}{\sqrt{2\pi N}(N/e)^N}\nonumber\\&\approx \frac{1}{2} e^{-2\xi_{\kappa} M}\frac{(2\xi_{\kappa} M)^{2\xi_{\kappa} M/\epsilon}}{\sqrt{4\pi \xi_{\kappa} M/\epsilon}(2\xi_{\kappa} M/\epsilon e)^{2\xi_{\kappa} M/\epsilon}}.
\label{pny}
\end{align}
The approximation in the first line holds because $2\xi_{\kappa} M/N\gg 1$. The final-$M$ error exponent
\begin{align}
\tilde{r}_{\rm C\veryshortrightarrow D}^{\rm ASY}&(M)=-\frac{1}{M}{\rm ln}P_{\rm C\veryshortrightarrow D}^{\rm ASY}\nonumber\\&\approx\frac{1}{M}\Big[(1-\frac{{\rm ln}e\epsilon}{\epsilon})2\xi_{\kappa} M+
\frac{1}{2}{\rm ln}2\xi_{\kappa} M+{\rm ln}2\sqrt{2\pi/\epsilon}\Big].
\label{asy1}
\end{align}

%\begin{figure}[t]
%    \centering
%    \includegraphics[width=\linewidth]{FIG83.pdf}
%    \caption{The error performance for the Rayleigh-fading model. The parameters are $N_{{\rm S}}=0.001$, $N_{\rm E}=20$ and $\bar{\kappa}=0.01$. (a) Comparison of the error probability for the decision threshold $N=0, 1$ calculated by Eq.~(\ref{EPN}) (gray and dotted lines) and the method of Monte Carlo simulation (the round and square points). The two results are consistent. (b) Comparison of the  achievable error performance given by Eq.~(\ref{heli}) (red), the lower bound given by Eq.~(\ref{LB}) (purple, dashed), and the error performance of SFG receiver. }
%    \label{fig8}
%\end{figure}

%\renewcommand\refname{Reference}
%\bibliographystyle{abbrv} % or try abbrvnat or unsrtnat
%\bibliography{myrefbib} % refers to example.bib

\end{document}